\documentclass{natureprintstyle}
\usepackage{epsfig,caption}
\usepackage{color}
\usepackage{bm}
\usepackage{graphicx}
\usepackage{longtable}
\usepackage{amssymb}
\usepackage{rotating}

\usepackage{hyperref}

\def\msun{\,M_{\odot}}

\title{Ubiquitous Time Variability of Integrated Stellar Populations}

\author{Charlie Conroy$^1$, Pieter G. van Dokkum$^2$ \& Jieun Choi$^1$}

\begin{document}

\maketitle

\let\thefootnote\relax\footnote{

\begin{affiliations}
\item Department of Astronomy, Harvard University,  Cambridge, MA, USA

\item Astronomy Department, Yale University,  New Haven, CT, USA
  
\end{affiliations}
}

\vspace{-3.5mm}
\begin{abstract}

Long period variable stars arise in the final stages of the
  asymptotic giant branch phase of stellar evolution. They have
  periods of up to $\sim1000$d and amplitudes that can exceed a factor
  of three in the $I$-band flux.  These stars pulsate predominantly in
  their fundamental mode\cite{Fox82,Wood96,Ita04}, which is a function
  of mass and radius, and so the pulsation periods are sensitive to
  the age of the underlying stellar population\cite{Vassiliadis93}.
  The overall number of long period variables in a population is
  directly related to their lifetime, which is difficult to predict
  from first principles because of uncertainties associated with
  stellar mass-loss and convective mixing.  The time variability of
  these stars has not been previously taken into account when modeling
  the spectral energy distributions of galaxies.  Here we construct
  time-dependent stellar population models that include the effects of
  long period variable stars, and report the ubiquitous detection of
  this expected `pixel shimmer' in the massive metal-rich galaxy
  M87. The pixel light curves display a variety of behaviors,
  including linearly rising and falling curves, semi-periodic curves,
  and sudden increases or decreases in the flux level.  The observed
  variation of $0.1-1$\% is very well matched to the predictions of
  our models.  The data provide a strong and novel constraint on the
  properties of variable stars in an old and metal-rich stellar
  population, and we infer that the lifetime of long period variables
  in M87 is shorter by approximately 30\% compared to predictions from
  the latest stellar evolution models.

\end{abstract}

In typical massive galaxies with $\sim10^{11}$ stars, the variation in
the total light due to long period variables will be small as the
summed light curves of many such stars effectively cancel each other
out (with random phases the net effect scales as $N^{-1/2}$).  If the
light is spread out over many (e.g., $\sim10^4-10^6$) pixels, then the
number of stars per pixel can range from $10^4-10^7$ and in this
regime the number of asymptotic giant branch stars per pixel is small
and governed by Poisson statistics.  This `semi-resolved' regime is
well known\cite{Tonry88} and the expected surface brightness
fluctuations due to Poisson statistics of rare luminous stars have
been observed and studied in several hundred nearby
galaxies\cite{Tonry01,Blakeslee09,vanDokkum14}.  We expect in this
regime to be able to detect the presence of variable stars through the
time dependence of the pixel flux (i.e., the pixel light curve):
essentially every pixel is expected to `shimmer' on timescales of
several hundred days.

In order to quantify the expected pixel shimmer, we created a stellar
population model at solar metallicity that includes the time-dependent
effect of long period variables.  We started with a new library of
stellar isochrones (Choi et al., in prep) that densely samples fast
phases of stellar evolution, and assigned periods to evolved stars
assuming that they pulsate in the fundamental
mode\cite{Vassiliadis93}.  We then used observations of variable stars
in the Galactic bulge from the OGLE survey to estimate a
period-amplitude relation in the
$I-$band\cite{Groenewegen05,Soszynski09,Soszynski13}.  A smooth
surface brightness model of the giant elliptical galaxy M87 was used
to specify the luminosity within pixels of size $0.2''\times0.2"$. The
pixel luminosity was used to normalize the weights in the isochrone
assuming a Salpeter initial mass function\cite{Salpeter55}.  For each
pixel the number of giants was drawn from a Poisson distribution and
the time evolution of the flux for each giant was given by its
associated period and amplitude and initialized with a random phase.
An illustration of this time-dependent model for M87 is shown in
Fig. 1.  The variable star part of our model has a tunable parameter,
the long period variable star weight, which can be interpreted as the
typical lifetime of such stars.  Further details regarding the
modeling are provided in the Methods section.

We sampled the model with the same cadence and applied the same photon
counting uncertainties as existing observations of M87 (see below).
The resulting model pixel light curves are shown in Fig. 2 (blue
lines).  These pixels were selected to have peak-to-peak flux
variation $>1.5$\%.  While rising and falling curves are clearly seen,
one also sees that a $<100$ day observing window can by chance sample
a light curve at a phase that appears relatively flat.  A $>200$d
observing cadence would clearly be ideal for observing the effects of
long period variables in the integrated light of nearby galaxies.

To test these expected variations we analyzed archival data of the
galaxy M87 from the {\it Hubble Space Telescope} collected over 72
days in 2005 \cite{Waters09,Peng09,Bird10}.  Imaging was obtained in
both the F606W and F814W filters with the Advanced Camera for Surveys.
We focused our analysis on the F814W imaging as the F606W data were
generally of lower quality (both due to a shorter exposure time and
the fact that only a single exposure was obtained per visit, which
made it difficult to clean the images of blemishes such as hot pixels
and cosmic rays).  The data were processed via the standard {\it HST}
pipeline.  In total 52 separate images, each with a depth of 1440s,
were considered in this analysis.  Globular clusters in the field were
used to refine the astrometric alignment with subpixel shifts.
Accurate subtraction of the background was achieved with several
additional corrections to the standard {\it HST} pipeline, as detailed
in the Methods section.  Pixels that deviated by more than 30\% from a
smooth model of the light profile were masked.  This effectively
removed all visible globular clusters, background galaxies, the chip
gap, and edge effects.  We also masked the central region and the
well-known jet in M87.  The data were binned $4\times4$ to
$0.2''\times0.2''$ in order to reduce the spatial coherency imposed by
the PSF (note that the models were also spatially binned and have been
convolved by the PSF in order to emulate the observations as closely
as possible).

Example pixel light curves for M87 are shown in Fig. 3.  The error
bars represent photon counting uncertainties only; the solid line is a
5-point boxcar averaging of the data.  We detect coherent variation in
the pixel light curves that is qualitatively consistent with our model
expectations.  These examples were chosen to highlight the level of
variety seen in the data.  We note that a significant fraction of
pixel light curves show no evidence of variation within the noise
limits of the data.  We show below that this is expected if long
period variables are the source of the variation.

We have quantified the pixel light curves by fitting each curve with a
linear function; the best-fit slope and uncertainty were recorded.
The resulting distribution of slopes (in units of the uncertainty) is
shown in Fig. 4.  We find that 24\% of pixels (48,100 out of 202,000)
show $>2\sigma$ evidence for variation.  In our model there are on
average 1.5 long period variable stars responsible for variation in
each $>2\sigma$ detection.  This implies a statistical detection of
$\sim72,000$ variable stars in M87.  When averaged over the central
$1'\times1'$ field of view, the model predicts on average 0.5 variable
stars per pixel.  In Fig. 4 we compare the observations to the model
predictions.  We show the sensitivity of the model light curve
statistics to both the stellar population age and variable star
parameters, and also the posterior probability distributions that
result from fitting the model to the observed histogram when allowing
the age and relative variable star weight to vary (we do not include
the tails of the distribution in the fit as the data are slightly
asymmetric beyond $|$slope/err$|\approx7$).  The latter is an overall
factor controlling the contribution of variable stars to the
integrated light relative to the predictions of a stellar evolution
model (see the Methods section for details). The pixel light curves
provide a strong constraint on a combination of the age and variable
star weight.  The dashed line shows the best-fit age estimated from
modeling the integrated light spectrum of the central region of
M87\cite{Conroy12}, which allows us to break the degeneracy.  It is
noteworthy that the best-fit long period variable star weight is less
than one, suggesting that such stars in M87 may have shorter lifetimes
than current solar metallicity stellar evolution models predict (see
the Methods section for a discussion of the effects of metallicity).

This is not the first detection of time variability in the pixel
fluxes of nearby galaxies; previous work predicted the occurrence of a
gravitational microlensing signal at the pixel level\cite{Gould95},
which was subsequently observed\cite{Baltz04}.  Novae have also been
observed in nearby galaxies\cite{Ferrarese03}, and indeed we
identified $\approx15$ novae through visual inspection of the pixel
light curves for M87.  However, novae and microlensing events are rare
(though bright) events.  An important distinguishing feature of the
time variation caused by long period variable stars is the ubiquity
--- as Fig. 4 shows, 24\% of the pixels show $>2\sigma$ evidence for
variation.

There are relatively few constraints on the stellar evolutionary phase
that gives rise to long period variables.  The best constraints
to-date on this phase are confined to the Magellanic Clouds, which
have sub-solar metallicities characteristic of low mass galaxies.  The
observations reported here have provided a direct constraint on this
important stellar evolutionary phase in a massive, high metallicity
galaxy.  New stellar evolution models over-predict the lifetimes of
long period variables by approximately 30\% if one adopts a
spectroscopic age for M87 of 10 Gyr.  An older mean population age
would reduce the mild tension between the models and observations.
Constraints such as these on highly evolved, luminous stars are
essential for interpreting light from more distant, massive and
metal-rich galaxies across the universe.

The detection of time variation in the integrated light of nearby
galaxies opens the possibility to derive stellar population ages in
these systems by a completely different approach from conventional
techniques.  In the future, one could imagine high cadence
observations of nearby galaxies on $>100$d baselines to detect the
period distribution of long period variables by analyzing power
spectra of the time series data.  This technique is not limited to old
stellar systems; on the contrary, younger systems will show
considerably greater temporal variation.  For example, based on our
models, we expect that 4\%, 14\%, and 22\% of pixels with $10^5$ stars
will show $>1$\% absolute flux changes over 100d for ages of
$10^{10}$, $10^9$, and $10^8$ yr, respectively.  The larger effect at
younger ages is due primarily to the larger fractional contribution of
long period variables to the total light (see the Methods section for
details).  It will therefore be relatively straightforward to perform
similar studies on nearby spiral galaxies, where the signal will be
much stronger.  At a basic and fundamental level, each pixel of a
galaxy measurably varies in time, and this variation encodes unique
information on its underlying stellar population.

\begin{addendum}
\item [Acknowledgements] We thank Martin Groenewegen
    for useful discussions and the referees for comments that improved
    the quality and clarity of the manuscript.  C.C. thanks Brad
    Holden and Connie Rockosi for asking the question that provided
    the spark for this paper: ``Can one detect Mira variables in
    integrated light?''

\item[Author Contributions] C.C. constructed the
    models, led the data processing, and contributed to the analysis
    and interpretation.  P.v.D.\ contributed to the analysis and
    interpretation.  J.C. generated the stellar evolution models and
    contributed to the analysis and interpretation.

\item[Author  Information]  Reprints and permissions
    information is available at npg.nature.com/reprintsandpermissions.
    Correspondence and requests for materials should be addressed to
    C.C.\ (cconroy@cfa.harvard.edu).

\end{addendum}

\newpage
\begin{figure*}
\centerline{
\includegraphics[width=0.9\textwidth]{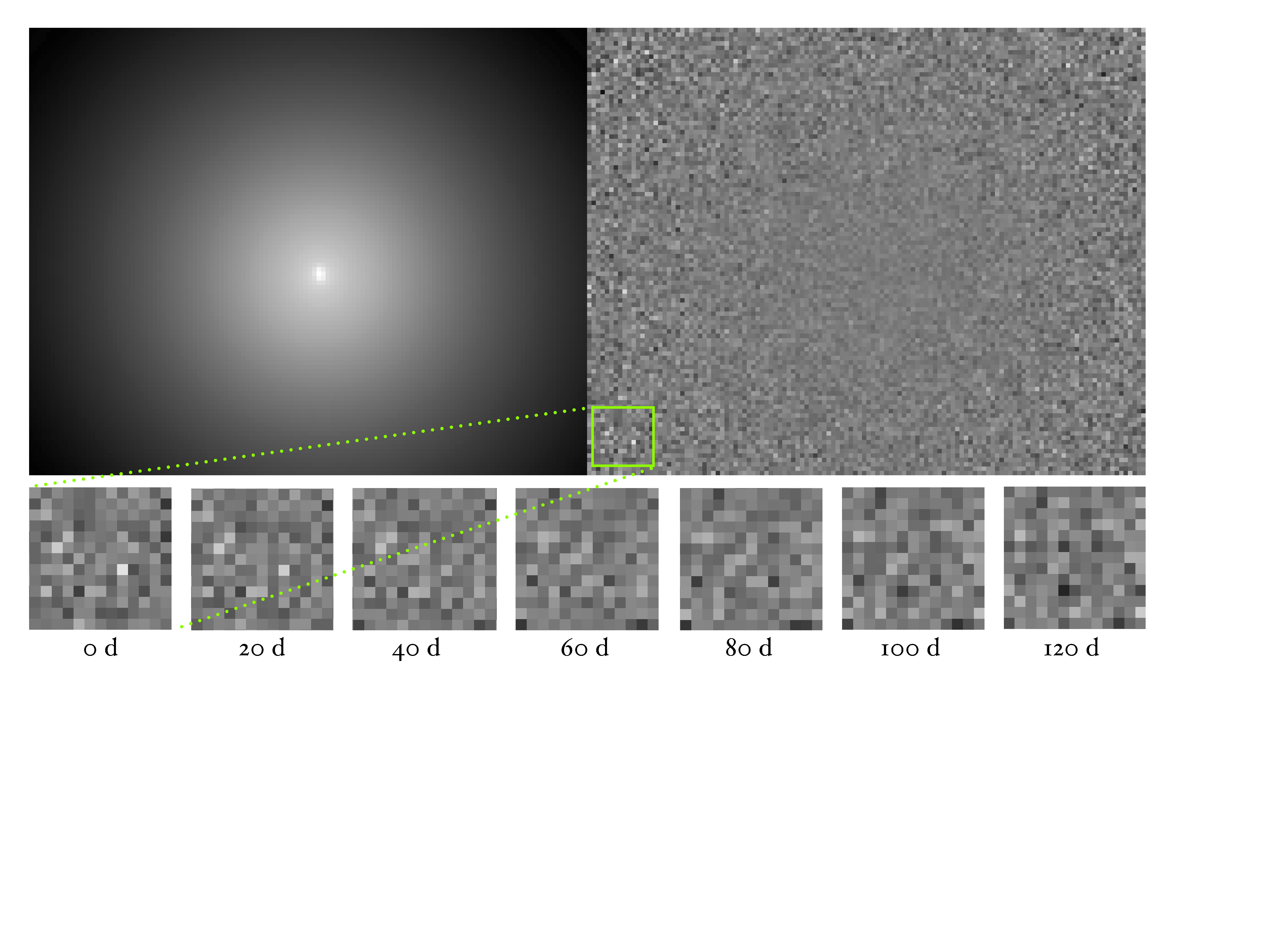}
}
\vspace{-4mm}
\caption{\textbf{Illustration of pixel shimmer.}  Model prediction of
  the effect of long period variables on integrated light.  Upper
  Left: A smooth model for the surface brightness profile of M87.
  Upper Right: The flux at $t=0$ divided by the mean flux over 1000d
  within each pixel.  Lower panels: Zoom-in on the lower left corner,
  showing snapshots at 20d intervals.  Notice the coherent variation
  in brightness of individual pixels. }
\vspace{-4mm}
\end{figure*}

\begin{figure*}
\centerline{
\includegraphics[width=0.6\textwidth]{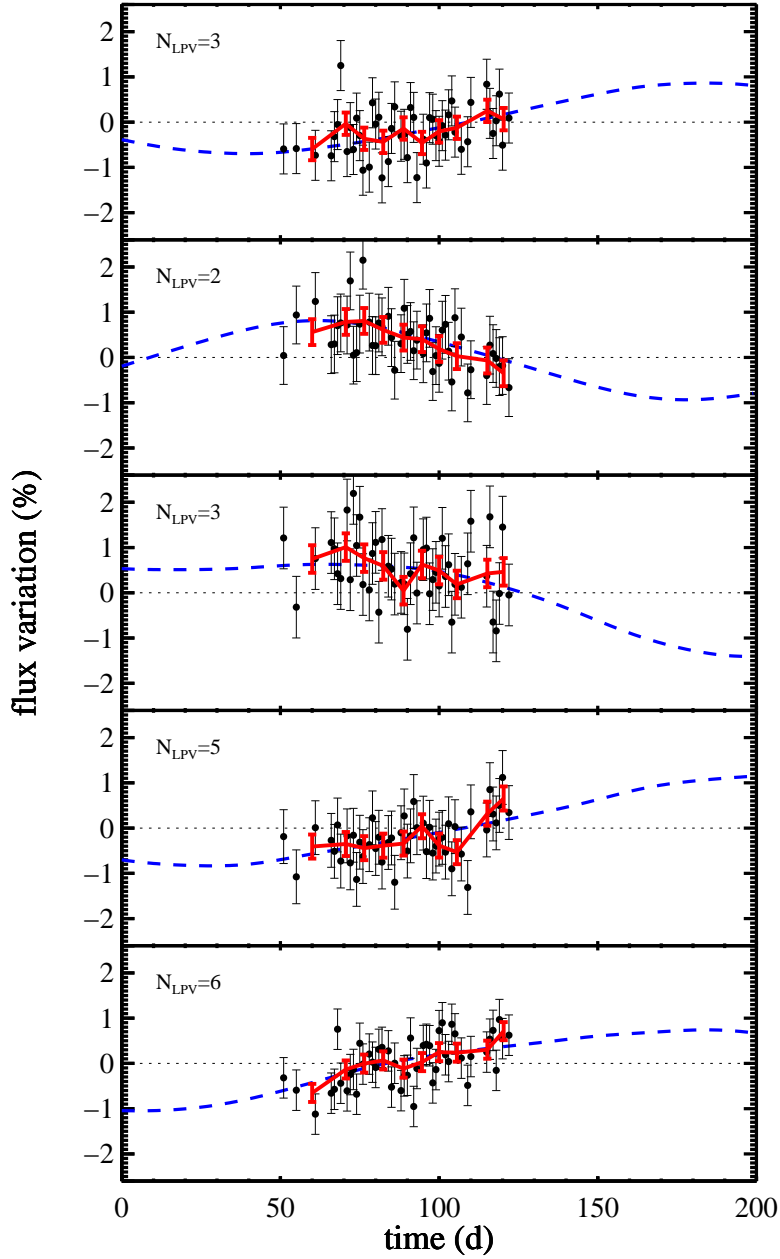}
}
\vspace{-4mm}
\caption{\textbf{Simulation of pixel light curves.}  Modeled relative
  flux variation over 200 days for five random pixels selected to have
  a peak-to-peak flux variation of $>1.5$\%.  The underlying stellar
  population is old and metal-rich and includes long period variable
  stars.  The noise-free model (dashed blue lines) is compared to a
  simulation of the M87 data, including the photon counting noise
  ($1\sigma$) and cadence of the observations (symbols), and a boxcar
  average of the simulated data (red lines).  In each panel the number
  of long period variable stars per pixel with periods $>150$d is
  indicated in the legend.}
\vspace{-4mm}
\end{figure*}

\begin{figure*}
\centerline{
\includegraphics[width=0.9\textwidth]{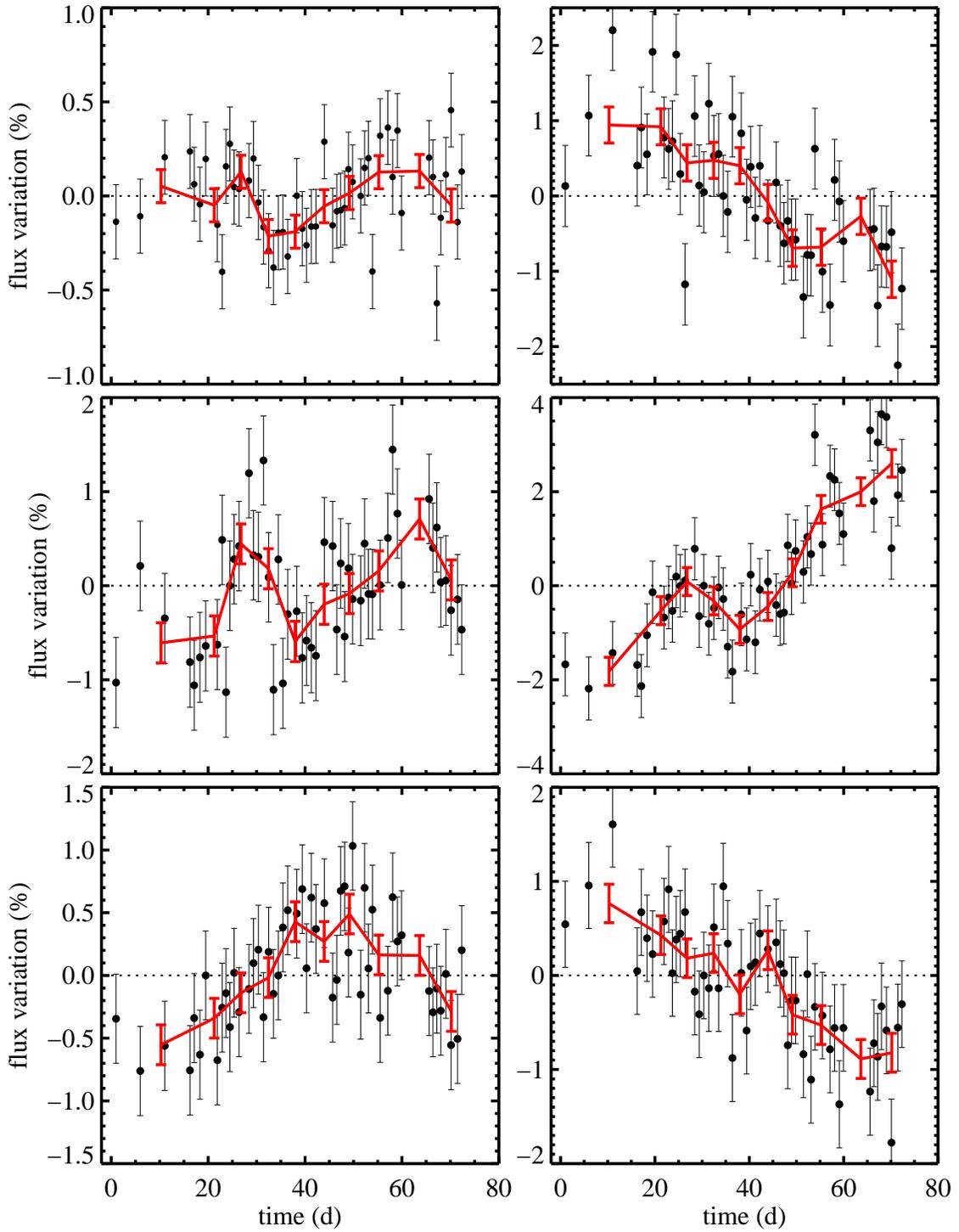}
}
\vspace{-4mm}
\caption{\textbf{Observed pixel light curves for M87.}  Relative flux
  variation over 72 days of pixels in M87.  We unambiguously detect
  the `pixel shimmer' due to the contribution of long period variables
  to the integrated light.  These pixels were selected to highlight
  the variety of morphology of the light curves, including rising,
  falling, periodic, and peculiar curves.  Errors represent $1\sigma$
  photon counting uncertainties.}
\vspace{-4mm}
\end{figure*}

\begin{figure*}
\centerline{
\includegraphics[width=0.9\textwidth]{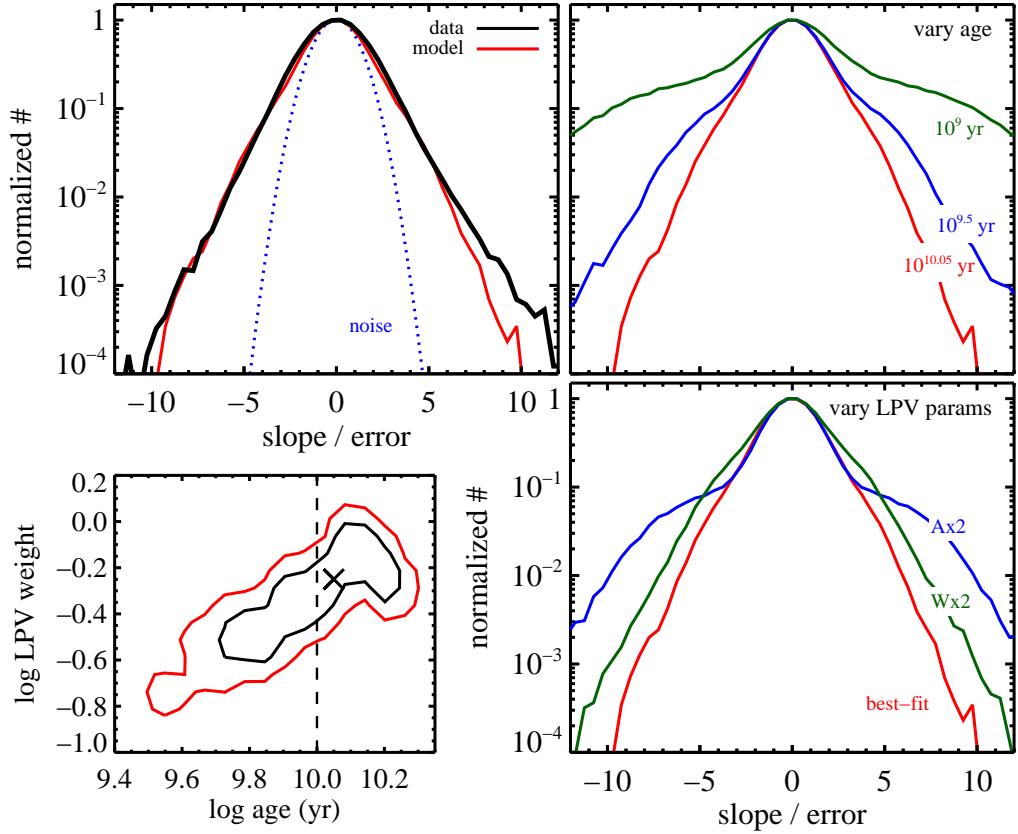}
}
\vspace{-4mm}
\caption{\textbf{Statistics of the pixel light curves.}  Distributions
  of the best-fit linear slope of the pixel light curve in units of
  the $1\sigma$ uncertainty on the slope.  The data are compared to
  several models including a variable star-free model (labeled
  ``noise'' in the upper left panel), and models varying the age,
  variable star amplitude, and weight.  Models with varying age and
  weight were fit to the observed histogram and the 1 and 2 sigma
  confidence limits on these parameters are shown in the lower left
  panel.  The cross symbol denotes the best-fit, shown as a red line
  in the other panels, and the vertical dashed line indicates the
  best-fit age from fitting the integrated light spectrum.  The lower
  right panel shows the effect of doubling the amplitudes of the long
  period variables (``Ax2''; blue line) and the weights (``Wx2'';
  green line).}
\vspace{-4mm}
\end{figure*}

\clearpage

\setcounter{page}{1}
\setcounter{figure}{0}
\setcounter{table}{0}
\renewcommand{\thefigure}{S\arabic{figure}}
\renewcommand{\thetable}{S\arabic{table}}


\begin{center}
{\bf \Large \uppercase{Methods} }
\end{center}

\noindent
{\bf Data reduction and tests.}  Owing to the small expected amplitude
of the time-dependent flux signal for M87, great care was taken to
control systematic effects.  In this section we describe the details
of the data reduction procedure and the additional corrections that
were applied to the images.

We began with the publicly available {\it HST} images, in which the
four dithered exposures per visit were combined and astrometrically
aligned, resampled by the drizzling process, and cosmic rays were
removed.  The public images include flat field corrections and a
standard sky subtraction.  We used five globular clusters to refine
the astrometric alignment via subpixel shifts (using bilinear
interpolation).  The mean shift was 0.25 pixels (in both x and y
directions) in the unbinned images.  All of our analysis was performed
on images binned $4\times4$, so these shifts are a tiny fraction of
the final pixel size.

Owing to the large angular extent of M87, the standard ACS pipeline is
not able to accurately measure the true sky background.  We therefore
applied a correction to the sky subtraction. We assumed that the true
M87 surface brightness profile is that reported by
Kormendy\cite{Kormendy09}, which was derived by combining a variety of
space and ground-based data.  Using this profile, we estimated the sky
background in our ACS images by minimizing the residuals between the
ACS data and Kormendy's profile with the sky background,
normalization, and a linear color gradient as free parameters (the
latter is to account for differences between our F814W filter and
Kormendy's $V$-band profile).  This was done separately for each of
the 52 images.  We refer to this as the primary sky background
correction.

In order to test the fidelity of the images over the 72 days we
selected three background galaxies and measured their fluxes within an
8 pixel aperture.  These galaxies should show no detectable temporal
variation.  The resulting temporal variation of the total flux from
these galaxies is shown in panel a of Extended Data Fig. 1.  There are
no obvious time-dependent trends.  However the scatter is 0.5\%, which
is relatively large compared to the signal of interest (of order 1\%).
We therefore made several additional modifications to the sky
background levels in an effort to reduce the scatter.

We identified three additional background galaxies (i.e., not the ones
used to measure the flux variation in Extended Data Fig. 1) and
measured their flux variation over the duration of the observations.
Under the assumption that these sources should have no intrinsic flux
variation, we determined a sky background correction necessary to
bring the flux of the background galaxies to a constant.  The average
correction determined this way was 0.002 ct/s.  At this point in the
analysis the distribution of pixel light curve slopes showed a slight
preference for positive slopes (the mean slope/error was +0.5).  Under
the assumption that the true distribution should have a mean of zero,
we subtracted a linearly varying sky background component (which
scaled as $5\times10^{-5}\,t$).  These two corrections yield a
distribution of pixel light curve slopes with zero mean (by
construction), and a temporal flux variation in the three reference
background galaxies with a scatter of 0.2\% as shown in panel b of
Extended Data Fig. 1.  Moreover, a 5-point boxcar average of the light
curve of the background galaxies shows flux variation at the
$\lesssim0.1$\% level.  From this test we conclude that it should be
possible to measure intrinsic flux variation at the sub percent level,
at least for pixels where photon counting noise is not the dominant
source of uncertainty.

We emphasize that the additional sky background corrections discussed
above do not materially change our conclusions.  While these
corrections result in a shift in the histogram of slope/err values,
they have no effect on the {\it width} of the distribution.  Moreover,
the example pixel light curves shown in Fig. 2 are unchanged within
their $1\sigma$ error bars.

Approximately half of the exposures were obtained at a detector
location offset by $60-70$ pixels compared to the other half of the
exposures.  This provides a further test that the trends shown in Fig.
3 and the statistics in Fig. 4 are not dominated by unknown systematics
at the level of the detector; if they were, one would have expected to
see flux variation that correlated with the dither pattern, but
such correlations, if present, are within the noise limits of the data.

As a final test of both the data reduction and our results, in
Extended Data Fig. 2 we show histograms of the flux variation over the
72 day observing window.  The flux variation was computed by
temporally binning the exposures by five to reduce the Poisson noise
in the measurement and computing (max-min)/mean flux at each pixel.
The results are shown for three bins of pixel fluxes (the legends show
the cuts in units of counts per second and the total number of pixels
per bin).  The data (black lines) are compared to our best-fit model
as derived from fitting the pixel light curve slopes (red lines) and a
model without long period variables (blue lines).  The good agreement
between the model and data in all three panels is a strong indication
that our measurements are reliable, as the panels probe a factor of 40
in dynamic range in pixel fluxes.  Systematic issues with e.g., the
sky subtraction would show up most strongly in the pixels with low
count rates, and yet the observations and models agree very well in
that regime.  Moreover, this flux variation metric is model
independent and so the difference between the variable star-free model
and the observations provides further strong support that the
variation detected in the observations is real and not an artefact of
some unknown systematics.  There do exist subtle differences between
the model and data that vary as a function of the pixel flux, but this
could be due to changes in the underlying stellar populations as the
pixels with low fluxes are in the outskirts, where the ages and
metallicities of the stars are expected to differ from the central
regions.

\noindent
{\bf Modeling long period variables.}  Here we provide additional
details regarding the incorporation of long period variables in the
stellar population synthesis modeling.  We start with stellar
isochrones that include all relevant evolutionary phases, including
thermally-pulsating asymptotic giant branch (AGB) stars.  We include a
model for circumstellar dust around these stars, which results in
dimmer stars especially for the most intrinsically luminous and
evolved stars\cite{Villaume15}.  Periods (in days) are assigned
according to the following equation:
\noindent
\begin{equation}
{\rm log}\,P = -2.07 + 1.94\,{\rm log}(R/R_{\odot}) - 0.9\,{\rm log}(M/M_{\odot}),
\end{equation}
\noindent
which assumes that the stars pulsate in their fundamental
mode\cite{Vassiliadis93}.  Next, we require a relation between pulsation
period and amplitude.  This relation is shown in Extended Data Fig. 3
for stars in the Galactic bulge from OGLE data\cite{Soszynski13}.
Symbols are color-coded according to the type of pulsator.  The dashed
lines are the adopted period-amplitude sequences:
\begin{equation}
{\rm log}\,A = 0.5\, {\rm log}\, P-1.25,
\end{equation}
\noindent
for the Mira sequence (${\rm log}P>2.2$), and:
\begin{equation}
{\rm log}\,A = 2\, {\rm log}\, P-5,
\end{equation}
\noindent
for the semi-regular variable (SRV) sequence ($1.0<{\rm log}P<2.2$).
In the equations above $P$ is in days and the amplitude $A$ is in the
$I-$band in magnitudes.  We note that the SRV sequence is included for
completeness but has a very small effect on the model predictions.

The equations above, along with the initial mass function weights
determined by the masses of the AGB stars in the isochrones,
completely specify our default variable star model.  In order to
convert fluxes to luminosities we have assumed a distance to M87 of
16.7 Mpc\cite{Blakeslee09}.  In order to explore the constraining
power of the data, we considered variation in both the amplitude of the
long period variables, implemented as an overall scaling of all the
amplitudes by the same factor, and the weight given to the variable
stars in the population synthesis.  The latter can be interpreted as a
change to the typical lifetime.

We have taken great care to ensure that the long period variable phase
is well-resolved in the isochrone tables.  The isochrones were
constructed from 185 individual mass models and with 600 equivalent
evolutionary points in the thermally-pulsating AGB phase alone.  We
have run a variety of tests to ensure that our model predictions are
``converged''; for example we have created models with fewer
evolutionary points and fewer input mass models and the resulting
predictions are very similar.  For context, at 10 Gyr our isochrones
contain 350 points on the AGB with periods $>200$d, while the
publically available Padova\cite{Marigo08} isochrones contain only 3
such points.

Extended Data Fig. 4 quantifies the fractional contribution of long
period variables to the total flux of a stellar population as a
function of wavelength, age, and metallicity ([Z/H]).  The flux
contribution peaks in the age range of $10^{8.5}-10^9$ yr and
increases toward redder bands.  The trend with wavelength is a
reflection of the fact that variable stars are cool and so emit most
of their light in the near-IR.  We caution that the
wavelength-dependence shown here does not directly translate into the
wavelength-dependence of the time-dependent signal because the
period-amplitude relation also depends on wavelength.  As the
pulsation directly affects the radius and hence the temperature, for
these cool stars one expects and indeed observes that the amplitudes
are larger in the bluer wavebands\cite{Smith02}.  The
metallicity-dependence is relatively modest, at least over the range
[Z/H]$=-0.3$ to [Z/H]=$+0.3$, typical of massive galaxies.  It is
difficult to provide a simple explanation of the model metallicity
variation, as it depends not only on the variable star lifetime,
luminosity, and temperature, but also on the properties of the
underlying stellar population.

We do not expect metallicity to play a critical role in the
interpretation of the observations for several reasons.  First, as
noted in the previous paragraph, the models suggest a relatively weak
metallicity-dependence of the long period variable flux contribution.
Second, M87 harbors a metallicity gradient\cite{Kuntschner10},
extending from slightly super-solar in the inner $R_e/8$ to slightly
sub-solar at $R_e$, where $R_e$ is the effective radius.  Despite this
metallicity gradient, our best-fit model provides an equally good fit
to the pixel shimmer statistics in both the central region and the
outskirts, as shown in Extended Data Fig. 2.

We note here that individual long period variables have been detected
in nearby galaxies including the Magellanic Clouds\cite{Wood83},
M31\cite{Fliri06}, and M32\cite{Davidge04}.  The most distant galaxy
with secure detections of individual long period variable stars is NGC
5128\cite{Rejkuba03}, and in this case the observations were confined
to the outskirts where the stellar density was sufficiently low to
permit the separation of the brightest evolved stars from the
background sea of lower luminosity stars.  These observations of
individual long period variables should provide very useful
constraints on the modeling of such stars and we intend to make use of
these constraints in future work.

\noindent 
{\bf Trends with radius.}  The {\it HST} field of view covers
the central $3.3'\times 3.3'$ of M87, of which the inner
$\approx1'\times1'$ has $S/N\gtrsim100$ per pixel for the observations
that were analyzed herein.  Kormendy\cite{Kormendy09} reports an
effective radius of $R_e=3.2'$ so the region of the images with high
S/N covers the inner $\approx0.3R_e$.  Extended Data Fig. 5 shows
several important quantities as a function of $R/R_e$ for our best-fit
model of M87.  Panel a shows the stellar mass per pixel for the
underlying smooth stellar distribution.  Panel b shows the fraction of
pixels with $|$slope/error$|>$2.  In the main text we reported that
24\% of pixels reach this criterion, and in fact that percentage
remains approximately constant with radius.  The constancy is the
result of two opposing effects: at larger radius the number of stars
per pixel is lower, which implies a larger variable star signal.  The
effect on the slope scales approximately as $\sqrt{N}$ ($N$ being the
number of stars per pixel) as multiple variable stars with random
phases will cancel each other out in a central limit theorem-like
process.  However, at larger radius the S/N is lower, and for a fixed
exposure time this also scales as $\sqrt{N}$.  Thus, for a fixed
exposure time, the detectability of long period variables is fairly
constant with radius.

The lower two panels of Extended Data Fig. 5 show the model trends with
radius for a noise-free model (infinite S/N).  In this case it is
clear that the absolute effect of long period variables is larger at
larger radius.  Panel c shows the fraction of pixels with $>1$\%
peak-to-peak flux variation over 200 d.  Old stellar populations with a
pixel mass $<10^6\msun$ yield $>1$\% flux variation in $\sim10$\% of
the pixels.  Panel d compares the surface brightness fluctuation (SBF)
amplitude at a single epoch to the mean temporal variation over a 200
d baseline; the latter is smaller than the former by a factor of
$\approx5$.  The SBF amplitude is computed as the standard deviation
of the model flux divided by a smooth model for the flux.

We close by noting that while the overall effect of long period
variables on the integrated light is relatively modest at old stellar
ages, it is much more prominent for younger stellar populations, e.g.,
in the $10^8-10^9$ yr range.  Future work devoted to younger stellar
populations will therefore likely uncover a rich array of
observational signatures of time variable stellar populations.

\medskip \noindent
{\small \bf {Code Availability}} {\small {We have opted not to make
    the code used in this manuscript available because the data
    reduction and analysis is fairly straightforward and can be easily
    reproduced following the methods described herein.}}

\clearpage
\setcounter{page}{1}
\setcounter{figure}{0}
\setcounter{table}{0}
\captionsetup[figure]{labelformat=empty}

\renewcommand{\thefigure}{Extended Data \arabic{figure}}
\renewcommand{\thetable}{Extended Data \arabic{table}}


\begin{figure*}
\centerline{
\includegraphics[width=0.9\textwidth]{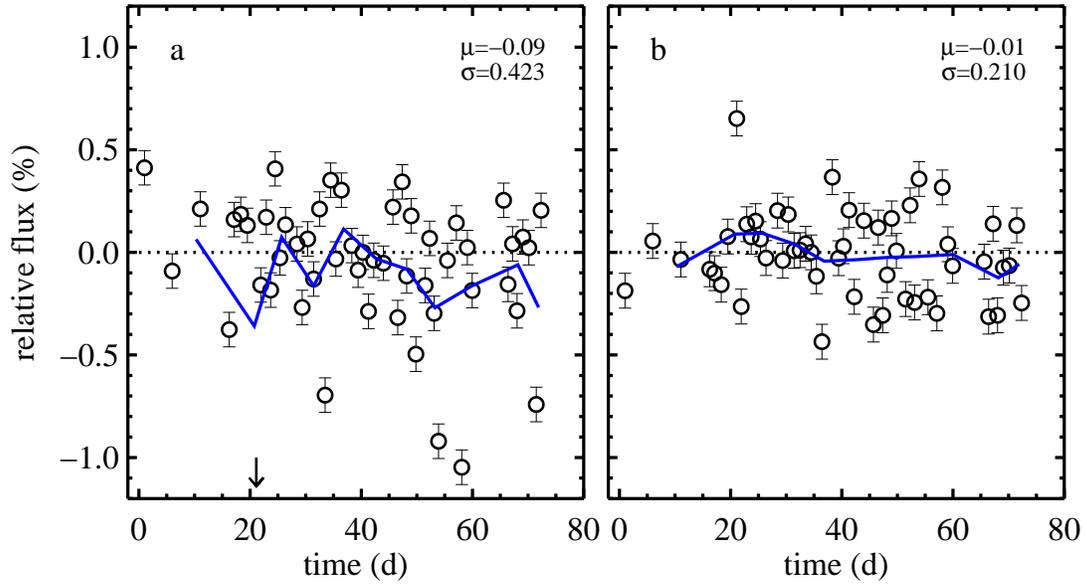}}
\vspace{-4mm}
\caption{\textbf{Extended Data Figure 1 $|$ Flux of background galaxies.}  Time variation of the flux of three background galaxies.  The
  background galaxies should show no intrinsic time variation in their
  flux and therefore serve as a test of the stability of the data.
  The mean and standard deviation are reported in each panel.  The
  $1\sigma$ error on each point due to photon counting uncertainty is
  0.09\%.  The solid line is a 5-point boxcar average.  {\bf a,} Flux
  variation after the standard data reduction including the primary
  sky background correction.  The arrow indicates a point that lies at
  -2.1.  {\bf b,} Flux variation after additional corrections were
  applied to the sky background levels.  These additional corrections
  allow us to achieve a stability of $\approx0.1$\% for boxcar
  averaged time series data. }
\vspace{-4mm}
\end{figure*}

\begin{figure*}
\centerline{
\includegraphics[width=0.9\textwidth]{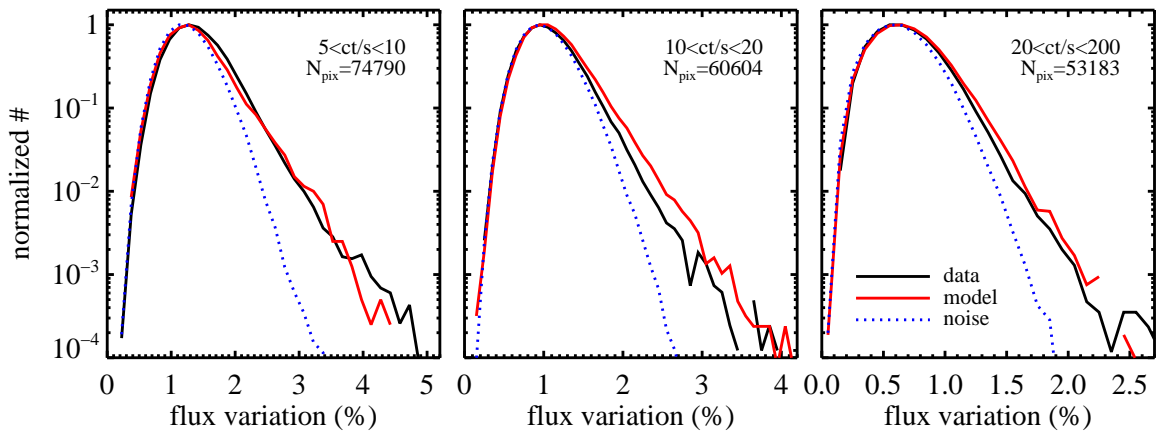}}
\vspace{-4mm}
\caption{\textbf{Extended Data Figure 2 $|$ Flux variation distributions.}  Distribution of (max-min)/mean fluxes over the 72 observing
  window, separated into three flux bins.   The data (black lines) are
compared to the best-fit model (red lines) and a noise-only model
(blue lines). }
\vspace{-4mm}
\end{figure*}

\begin{figure*}
\centerline{
\includegraphics[width=0.8\textwidth]{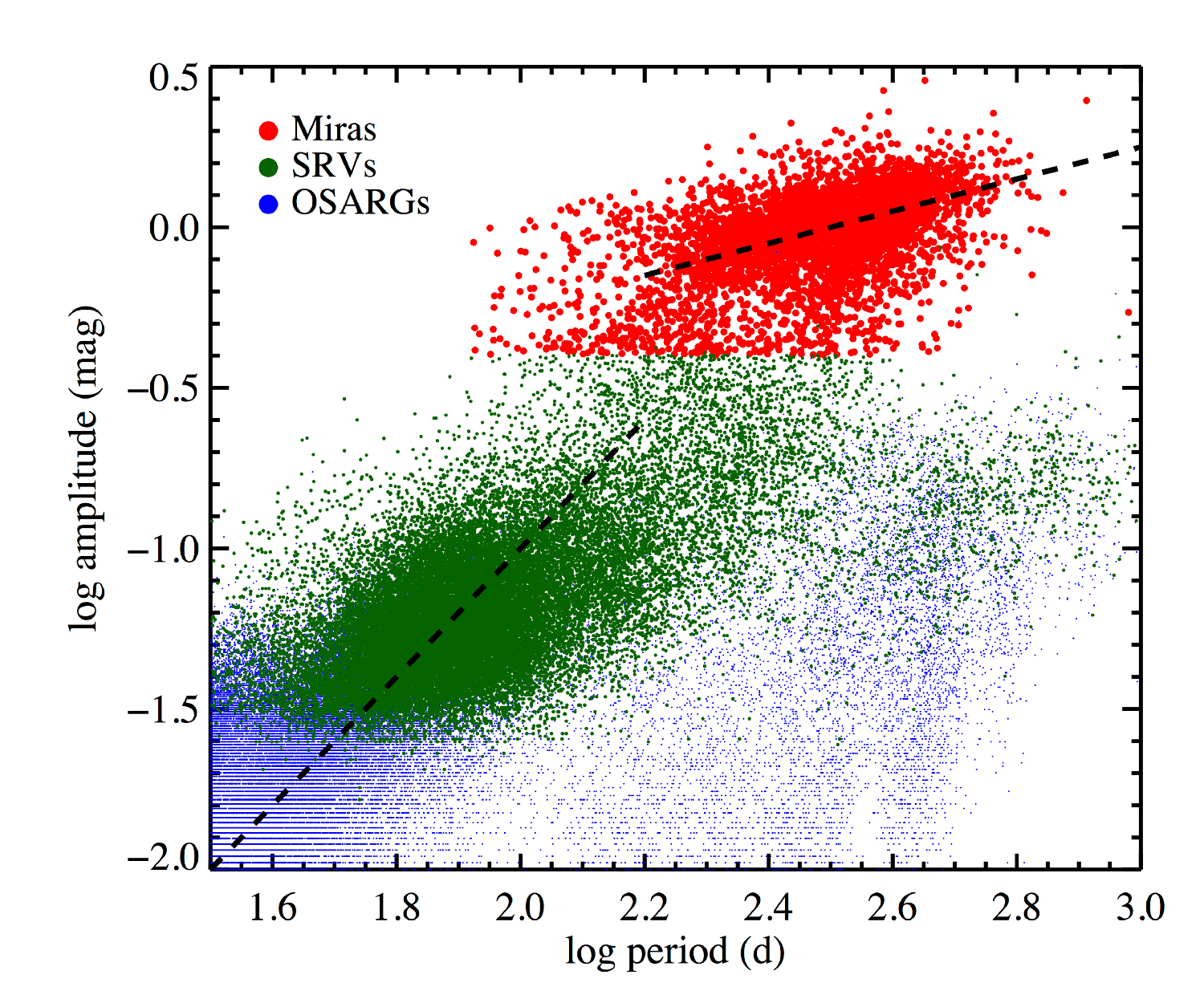}}
\vspace{-4mm}
\caption{\textbf{Extended Data Figure 3 $|$ Amplitude vs. period for
    luminous variable stars.}  Data are for Galactic bulge stars from
  the OGLE survey\cite{Soszynski13} measured in the $I-$band.  Lines
  are the adopted sequences for Miras and semi-regular variables
  (SRVs); these relations are used to assign pulsation amplitudes in
  our model.  OGLE small amplitude red giants (OSARGs) are shown for
  completeness. }
\vspace{-4mm}
\end{figure*}

\begin{figure*}
\centerline{
\includegraphics[width=0.9\textwidth]{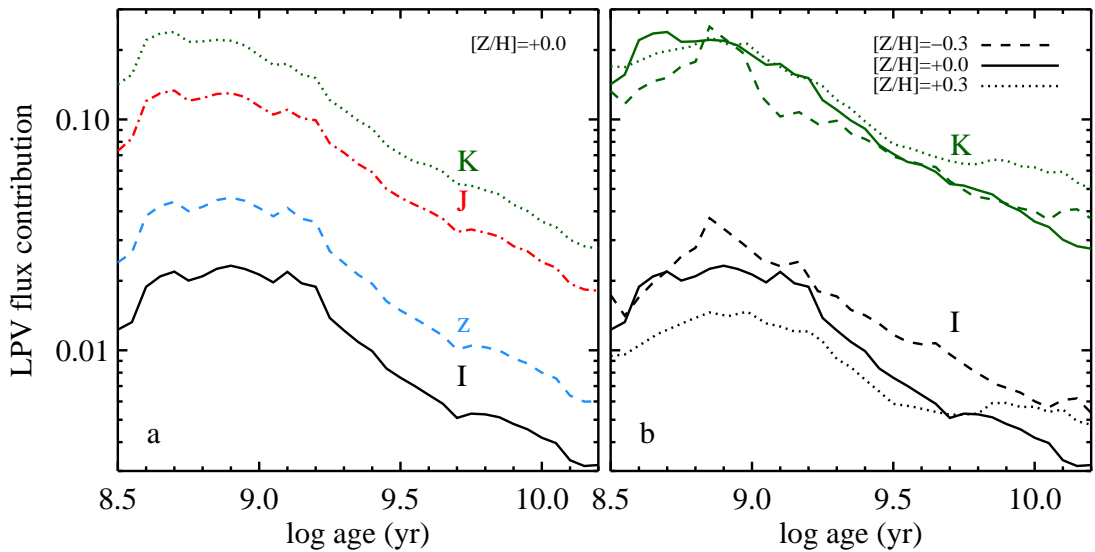}}
\vspace{-4mm}
\caption{\textbf{Extended Data Figure 4 $|$ Long period variable star flux contribution
      versus age, wavelength, and metallicity.}  {\bf a,}
  Fractional contribution to the total luminosity vs. age in four
  bandpasses from $I$ ($0.8\mu m$) through $K$ ($2.4\mu m$).  The flux
  contribution scales approximately as $t^{-1/2}$.  {\bf b,} Flux
  contribution vs. age and metallicity for the $I$ and $K$ bandpasses.}
\vspace{-4mm}
\end{figure*}

\begin{figure*}
\centerline{
\includegraphics[width=0.7\textwidth]{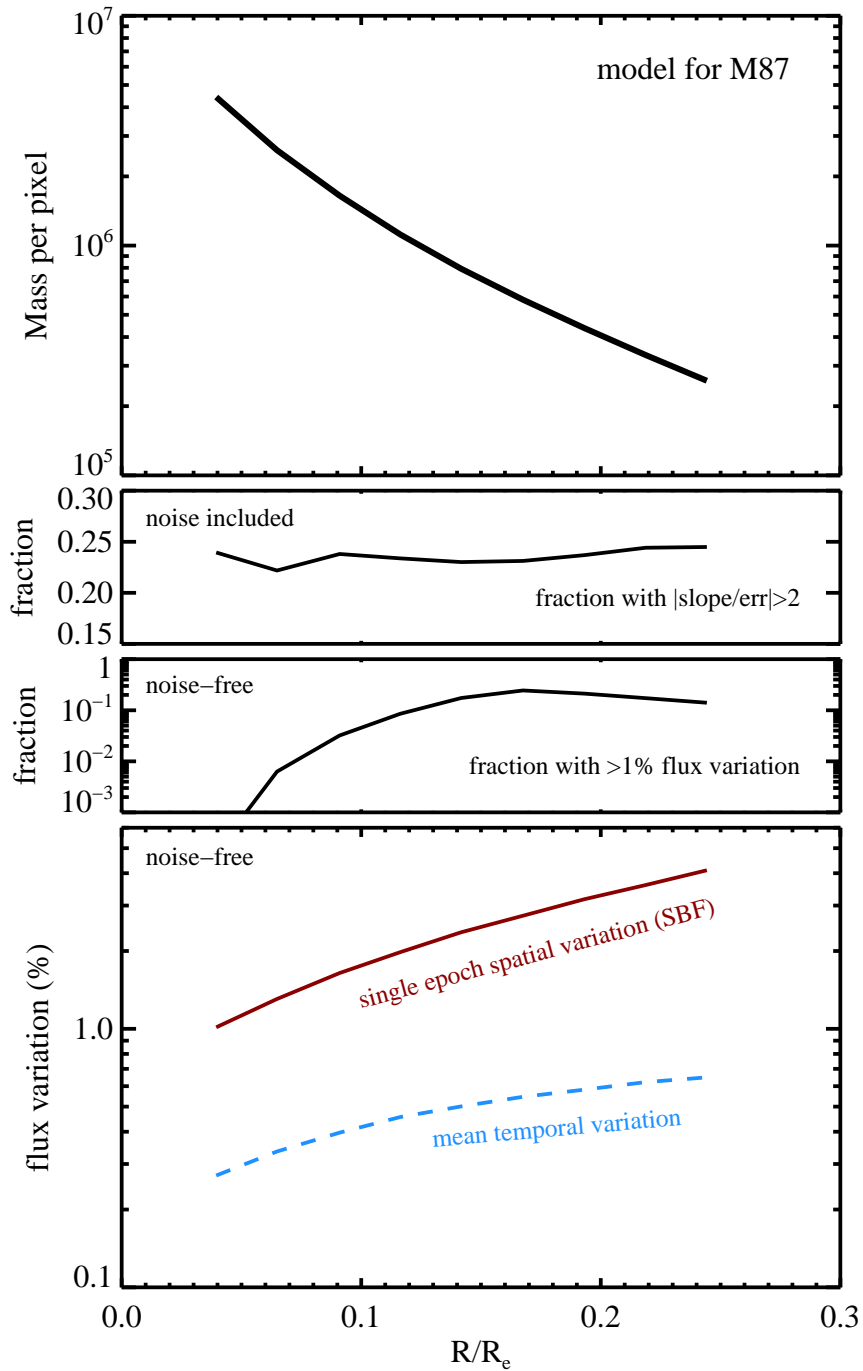}}
\vspace{-4mm}
\caption{\textbf{Extended Data Figure 5 $|$ Radial variation of model properties for M87.}  {\bf a,} Stellar mass per pixel for the smooth underlying
  model for M87 as a function of $R/R_e$ where $R_e$ is the effective
  radius.  {\bf b,} Fraction of pixels with $|$slope/err$|>2$.  {\bf
    c,} Fraction of pixels in a noise-free model with $>1$\%
  peak-to-peak flux variation over 200 days {\bf d,} Strength of SBF
  signal at a single epoch compared to the mean temporal variation due
  to variable stars in a noise-free model.  }
\vspace{-4mm}
\end{figure*}

\end{document}